\documentclass[%
 reprint,
superscriptaddress,
floatfix,
frontmatterverbose, 
nofootinbib,
 amsmath,amssymb,
 aps,
 prd,
longbibliography
]{revtex4-2}

\usepackage{cancel}
\usepackage{graphicx}% Include figure files
\usepackage{dcolumn}% Align table columns on decimal point
\usepackage{bm,ulem}% bold math
\usepackage[unicode=true,pdfusetitle,
 bookmarks=true,bookmarksnumbered=false,bookmarksopen=false,
 breaklinks=false,
 colorlinks=true,citecolor=steelblue]
 {hyperref}

\begin{document}

\definecolor{airforceblue}{rgb}{0.36, 0.54, 0.66}
\definecolor{steelblue}{rgb}{0.27, 0.51, 0.71}
\definecolor{amber}{rgb}{1.0, 0.49, 0.0}
%\preprint{APS/123-QED}

\title{Excited lepton triplet contribution to electroweak observables at one loop level}% Force line breaks with \\

\author{\textsc{M.~Rehman}}
\email{m.rehman@comsats.edu.pk}
\affiliation{Department of Physics, Comsats University Islamabad, 44000 Islamabad, Pakistan}
\author{\textsc{M.~E.~Gomez}}%
 \email{mario.gomez@dfa.uhu.es}
\affiliation{%
 Departamento de Ciencias Integradas, Facultad de Ciencias Experimentales, Campus El Carmen, Universidad de Huelva, Spain
}%


\author{\textsc{O.~Panella}}
\email{orlando.panella@cern.ch}
\affiliation{
 INFN, Sezione di Perugia, Via A. Pascoli, I-06123, Italy
}%
%\affiliation{
%\affiliation{%
% This line break forced with %\textbackslash\textbackslash
%}%


\date{\today}% It is always \today, today,

\begin{abstract}
In this paper, we present the one-loop radiative corrections to the electroweak precision observable $\Delta \rho$ coming from the $I_W=1$  multiplet excited leptons. We have calculated the couplings of the exotic lepton triplet to the vector bosons and ordinary leptons using effective Lagrangian approach. These couplings are then used to estimate the excited lepton triplet contribution to the $\Delta \rho$ parameter. The mass degenerate excited lepton contribution to $\Delta \rho $ is small and can be neglected. However, if the excited leptons are non-degenerate, their contribution can be large which can result in more stringent constraints on the excited fermion parameter space compared to the constraints from present experimental searches and perturbative unitarity condition.  

%\begin{description}
%\end{description}
\end{abstract}

\maketitle

%\tableofcontents

\section{\label{sec:level1}Introduction}

The Standard Model (SM) of particle physics has shown tremendous agreement with the experimental results however there are still some question that can not be answered by SM. For example SM can not explain why there are three generations of fermions and why fermion masses are the way there are. These kind of questions can be answered if one assumes the composite structure of the fermions. This assumption would require the SM to be a limiting case of a more fundamental theory vaild up to some high energy scale, called the composite scale $\Lambda$. Compositeness would predict the existence of heavy excited particles, of mass $M$, for each fermion state, however there is no satisfactory model, according to best of our knowledge,  that could reproduce the complete fermion spectrum. In the absence of the predictive model, we can use the effective Lagrangian approach to study the physics related to the compositeness.

There are already several efforts to study the physics at the composite scale. Most of them concentrated on the production of  excited fermions at colliders. For example production of excited states members of multiplets of isospin $I_W=0, 1/2$ was discussed in ~\cite{Baur:1989kv,Baur:1987ga}. In~\cite{PhysRevLett.105.161801,Aad:2016aa}, bounds on the masses of the excited fermions with $I_W=0,1/2$ were presented on the basis of experimental searches at LHC. On the phenomenological side, excited fermion contributions to the $Z$ pole observables were calculated in~\cite{GonzalezGarcia:1996nz,GonzalezGarcia:1996nz} for the isospin doublet states. However it has been shown that the higher isospin multiplets up to $I_W=1,3/2$ are also allowed by the standard model symmetries~\cite{Pancheri:1984sm}. This will result in the existence of exotic states such as quarks $U^+$ of charge $+5/3e$  and quarks $D^-$ of charge  $-4/3 e$. Phenomenological studies for these  exotic states were presented in ~\cite{Biondini:2012ny,Leonardi:2014epa,Biondini:2014dfa,Leonardi:2015qna}. Experimental studies have been able to put stringent bounds on the excited fermion masses ($M$) but so far there is no direct experimental evidence of the existance of these kinds of states. Phenomenological studies and experimental searches typically provide bounds in the parameter space $[\Lambda,M]$. In the absence of direct detection, indirect effects of excited fermions to the standard model observables can be a very good probe for the study of these states. 
Recently there have been interesting development in the study of the phenomenology of the effective interactions of excited fermions by computing unitarity bounds~\cite{Biondini:2019tcc} in the parameter space $[\Lambda, M]$. Such unitarity bounds have been shown to have a potential impact when compared with the bounds from the direct searches at colliders~\cite{Biondini:2019tcc}. 

In this paper, we study the indirect effects of excited fermions with isospin multiplets $I_W=1$, to the electroweak precision observables. Our main obejective is to derive bounds on the parameter space $[\Lambda, M]$ from the experimental   value of electro-weak precision observable $\Delta\rho$.
As a first step,  we considered the contribution of excited leptons since the ones from the quarks are expected to be of the same order. We have calculated the couplings of the excited fermions (leptons) of the triplet to the excited and ordinary fermions (leptons) using an effective field theory {lagrangian} approach which will be presented in section~\ref{model_setup}. The analytical results for the excited fermion (lepton) contribution to the $\Delta \rho$ observable  will be given in section~\ref{Aresults}. We will present our numerical analysis in section~\ref{Nresults}. Our conclusions can be found in section~\ref{sec:conclusions}. 

\section{Model set-up}
\label{model_setup}
The success of the Standard Model raised the hope of extending the particle spectrum beyond the known three families using weak isospin as was previously done for the case of Baryons and Mesons. It is believed that weak isospin can help us predict hypothetical new fermions without reference to the specific dynamic model of their building blocks (so called preons). Using $I_W=0, 1/2, 1, 3/2$, one can construct new states of fermions which can interact with the ordinary fermions via electroweak boson ($B^{\mu},W^{\mu}$) fields and gluon $G^{\mu a}$ field. In Table \ref{Lep-Multiplets}, we show all such states which can result from higher isospin lepton multiplets. For the sake of simplicity, we only list first generation of these excited states including only the leptonic sector. 

\begin{table*}[t]
\caption[Lepton multiplets]{\label{Lep-Multiplets}Lepton multiplets for $I_W=0,1/2,1,3/2$
, their charge $Q$, hypercharge $Y$ and the fields through which they couple to ordinary leptons.}
%\vspace{-0.45cm}
%\begin{center}
%\vspace{1 cm}
\begin{ruledtabular}\begin{tabular}{cccccl}%\hline
$I_{W}$ & Multiplet & $Q$ & $Y$ & Couple to & Couple through\\\hline
$0$ & $%
\begin{pmatrix}
E^{-}%
\end{pmatrix}
$ & $%
\begin{array}
[c]{c}%
-1
\end{array}
$ & $%
\begin{array}
[c]{c}%
-2
\end{array}
$ & $e_{R}$ & $B^{\mu}$\\\hline
$\frac{1}{2}$ & $%
\begin{pmatrix}
E^{0}\\
E^{-}%
\end{pmatrix}
$ & $%
\begin{array}
[c]{c}%
0\\
1
\end{array}
$ & $%
\begin{array}
[c]{c}%
-1
\end{array}
$ & $%
\begin{pmatrix}
\nu_{e}\\
e
\end{pmatrix}
_{L}$ & $B^{\mu},W^{\mu}$\\\hline
$1$ & $%
\begin{pmatrix}
E^{0}\\
E^{-}\\
E^{--}%
\end{pmatrix}
$ & $%
\begin{array}
[c]{c}%
0\\
-1\\
-2
\end{array}
$ & $%
\begin{array}
[c]{c}%
-2
\end{array}
$ & $e_{R}$ & $W^{\mu}$\\\hline
$\frac{3}{2}$ & $%
\begin{pmatrix}
E^{+}\\
E^{0}\\
E^{-}\\
E^{--}%
\end{pmatrix}
$ & $%
\begin{array}
[c]{c}%
+1\\
0\\
-1\\
-2
\end{array}
$ & $%
\begin{array}
[c]{c}%
-1
\end{array}
$ & $%
\begin{pmatrix}
\nu_{e}\\
e
\end{pmatrix}
_{L}$ & $W^{\mu}$\\
\end{tabular}
\end{ruledtabular}
%\end{center}
\end{table*}
Most of the literature adressing the phenome\-no\-lo\-gy of the excited fermions is based on the assumption that they have spin 1/2 and weak iso-spin $I_W=1/2$. Here we would like to extend our discussion to the case of higher iso-spin multiplets ($I_W=1$)~\cite{Baur:1987ga, Baur:1989kv,Boudjema:1992em}. Since the masses of the excited states are expected to be in the TeV range (i.e. much larger than the masses of the Standard Model fermions) it is customarily assumed that they acquire their mass prior to $SU(2)\times U(1)$ breaking.  Thus both the left-handed and right-handed components of the excited leptons belong to the same weak isospin multiplets (and they have the same quantum numbers). Therefore the corresponding coupling to the gauge bosons will be vector-like. Following the same notation as in \cite{GonzalezGarcia:1996nz} the  coupling of the  excited leptons to the gauge bosons  is given 
by the $SU(2) \times U(1)$ invariant (and CP conserving), effective
Langragian:
\begin{eqnarray}
{\cal L}_{FF} &=& 
- \bar{\Psi}^\ast \Biggl[ \left(g \frac{\tau^i}{2} \gamma^\mu W^i_{\mu} + 
g' \frac{Y}{2}  \gamma^\mu B_\mu \right) \Biggr.\nonumber \\&\phantom{=}& + \Biggl.
\left(\frac{g \kappa_2}{2 \Lambda} \frac{\tau^i}{2} \sigma^{\mu\nu}
\partial_\mu W^i_{\nu} + 
\frac{g' \kappa_1}{2 \Lambda}  \frac{Y}{2} 
\sigma^{\mu\nu}\partial_\mu  B_\nu \right) 
\Biggr] \Psi^\ast\nonumber\\
\label{l:ee:0}
\end{eqnarray}
where we have denoted with $\Psi$ the excited multiplet whose particle content is described in table~\ref{Lep-Multiplets}, $g$ and $g'$ are the gauge coupling constants of
$SU(2)$ and $U(1)$ respectively and $\kappa_1, \kappa_2$ are dimensionless couplings.  The constant  $\Lambda$ appearing in the dimension-5 part of the lagrangian in Eq.~\ref{l:ee:0} is the compositeness scale. 
In terms of the physical gauge fields, this can be written as: 
\begin{equation}
{\cal L}_{FF} = - \!\!\! \sum_{V=\gamma,Z,W} \bar{F} 
(A_{VFF} \gamma^\mu V_\mu + K_{VFF} \sigma^{\mu\nu} \partial_\mu V_\nu) F\; .
\label{l:ee}
\end{equation}
where $F$ denotes a generic excited fermion field appearing in the multiplet in table~\ref{Lep-Multiplets}.
Since we have assumed that the left- and right-handed excited
leptons have the same quantum numbers under the standard gauge
group, the dimension-four piece in Eq.~(\ref{l:ee}) is taken
vector-like.
Here we write down explicitly the couplings of the $I_W=0,1/2,1,3/2$ states of the multiplets in table \ref{Lep-Multiplets}. The higher multiplets of course include states with exotic charge like doubly charged leptons which were not included in the study in  \cite{GonzalezGarcia:1996nz} where only the doublet was considered.
The couplings $A_{VFF}$ are given by:
\begin{equation}
\begin{array}{ll}
A_{\gamma E^{-}E^{-}} =  - e &
\; , \;\;\;\;
A_{\gamma E^{0}E^{0}} =   0 \\
A_{\gamma E^{--}E^{--}} = -2e  &
\; , \;\;\;\; 
A_{ZE^{0}E^{0}} =  \displaystyle\frac{e}{ s_W  c_W}
\\
A_{ZE^{-}E^{-}} = \displaystyle  \frac{ e s_W }{ c_W} &
\; , \;\;\;\; 
A_{ZE^{--}E^{--}} = \displaystyle\frac{ -e (1- 2 s^2_W) }{s_W  c_W} \\ 
A_{WE^{0}E^{-}} = \displaystyle\frac{ e}{ s_W}  &
\; , \;\;\;\; 
A_{WE^{-}E^{--}} =  \displaystyle\frac{e }{s_W}\\
A_{W E^{0}E^{--}} = 0  &
\end{array}
\label{AV}
\end{equation}
while the couplings  $K_{VFF}$ are given by
\begin{equation}
\begin{array}{ll}
K_{\gamma E^{0}E^{0}} =
- \frac{\displaystyle e}{\displaystyle 2 \Lambda} (\kappa_2 - \kappa_1) &
,\;
K_{\gamma E^{-}E^{-}} =  
\frac{\displaystyle e}{\displaystyle 2 \Lambda} \kappa_1 \\
K_{\gamma E^{--}E^{--}} = 
- \frac{\displaystyle e}{\displaystyle 2 \Lambda}
(\kappa_2 + \kappa_1)  &
,\;
K_{WE^{0}E^{-}} =  
\frac{\displaystyle e \kappa_2}{\displaystyle 2 \Lambda s_W} \\
K_{WE^{-}E^{--}}= 
\frac{\displaystyle e}{\displaystyle 2 \Lambda s_W} &
,\;
K_{Z E^{-}E^{-}} = 
 \frac{\displaystyle e \kappa_1 s_W}{\displaystyle 2 \Lambda c_W} \\ 
 K_{Z E^{0}E^{0}} =  
\frac{\displaystyle e ( \kappa_1 s^2_W +  \kappa_2 c^2_W)}{\displaystyle 2 \Lambda c_W s_W} & \;,\\
K_{Z E^{--}E^{--}} =  
\frac{\displaystyle e ( \kappa_1 s^2_W -  \kappa_2 c^2_W)}{\displaystyle 2 \Lambda c_W s_W} 
\end{array}
\label{KV}
\end{equation}
The $SU(2)\times U(1)$ invariant and CP conserving  dimension-five effective Lagrangian describing
the coupling of the excited fermions to the usual fermions, which ensures the conservation of the electro-magnetic current is of the magnetic-type and 
  can be
written as \cite{Hagiwara:1985aa}
\begin{equation}
{\cal L}_{Ff} = - \frac{1}{2 \Lambda} {\bar \Psi^\ast} \sigma^{\mu\nu}
\left(g f_2 \frac{\tau^i}{2} W^i_{\mu\nu} + 
      g' f_1 \frac{Y}{2} B_{\mu\nu}\right) \psi_L 
+ \; \text{h. c.} ,
\label{l:eu:0}
\end{equation}
where $f_2$ and $f_1$ are dimensionless  factors of order unity  associated to the
$SU(2)$ and $U(1)$ coupling constants, and  $\sigma_{\mu\nu} = (i/2)[\gamma_\mu,
\gamma_\nu]$.  At tree-level they can  be
expressed in terms  of the electric charge, $e$, and the Weinberg
angle, $\theta_W$, as  $g=e/\sin\theta_W$ and
$g'=e/\cos\theta_W$. It is customary to  assume a pure left-handed structure
for the transition couplings in order to comply with  the strong bounds which are available
 from the measurement of the anomalous magnetic  moment of
leptons~\cite{Brodsky:1980aa,Renard:1982aa,Renard:1982ab}.
In terms of the physical fields, the Lagrangian (\ref{l:eu:0})
becomes
\begin{eqnarray}
{\cal L}_{Ff} &=& - \sum_{V=\gamma,Z,W} 
C_{VFf} \bar{F} \sigma^{\mu\nu} (1 - \gamma_5) f \partial_\mu V_\nu \nonumber \\
&\phantom{=}&- i \sum_{V=\gamma,Z} D_{VFf} \bar{F} \sigma^{\mu\nu} (1 - \gamma_5) f
W_\mu V_\nu +  \text{h.\! c.},
\label{l:eu}
\end{eqnarray}
where $F$ are the excited fermion states, $f$ the ordinary (SM) fermions  and $V=\gamma, Z,W$ are the physical vector boson fields. The non-abelian structure of
(\ref{l:eu:0}) introduces a  quartic contact interaction,
such as the second term in the r.h.s. of Eq.\ (\ref{l:eu}). In this
equation, we have omitted terms containing two $W$ bosons, which
do not play any role in our calculations.
Using \eqref{l:eu:0} and \eqref{l:eu}, we have calculated, for the case of $I_W=1$ lepton multiplets, the couplings $C_{VFf}$ and $D_{VFF}$ which are given in \eqref{CV} and \eqref{DV}. For the case of $I_W=0,1/2$ multiplets we refer the reader to \cite{GonzalezGarcia:1996nz}.
\begin{equation}
\begin{array}{ll}
C_{\gamma E^{-} e} =  - \frac{\displaystyle e }{\displaystyle \Lambda} f_{1}\ 
&
\;\; , \;\;\;\;
C_{Z E^{-} e} =  
- \frac{\displaystyle e c_W}{\displaystyle \Lambda s_W} f_{1}\ \\
C_{W E^{0} e} =  
\frac{\displaystyle e }{\displaystyle \Lambda s_W} f_1\ &
\;\; , \;\;\;\;
C_{W E^{--} e} =    
\frac{\displaystyle e }{\displaystyle \Lambda s_W} f_1\
\\
& 
\end{array}
\label{CV}
\end{equation}
and the quartic interaction coupling constant, $D_{VFf}$, is given by
\begin{equation}
\begin{array}{l}
D_{\gamma E^{0} e} = - D_{\gamma E^{--} e}=   
 \frac{\displaystyle -e^2 }{\displaystyle 4 \Lambda s_W } f_1\ \\
D_{Z E^{0} e} = - D_{Z E^{--} e} = 
 \frac{\displaystyle e^2 c_W}
{\displaystyle 4 \Lambda s^2_W } f_1\ \\
D_{W E^{-} e} =  
 \frac{\displaystyle - e^2 }
{\displaystyle 4 \Lambda s^2_W } f_1\
\end{array}
\label{DV}
\end{equation}

\section{Excited Lepton Contribution to \texorpdfstring{$\Delta \rho $}{Delta-rho}}
\label{Aresults}
\begin{figure}[htb!]
\begin{center}
\includegraphics[scale=0.10]{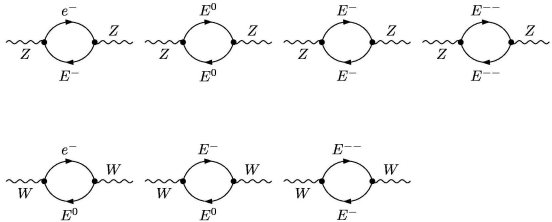}
\end{center}
\caption{$W$ and $Z$ boson self energy Feynman diagrams containing excited lepton in the loop.}
\label{fig:SE-FeynmanDiagrams}
\end{figure}

In this section we concentrate on the computation of the contribution of the excited leptons to $\Delta \rho$. This contribution can be considered as representative of the fermion contributions of the first family. Feynman diagrams for $Z$ and $W$ boson self energies involving ordinary leptons and excited leptons in the loop are shown in Fig.\ref{fig:SE-FeynmanDiagrams}.
The universal corrections to the electroweak precision observable $\Delta \rho$ can be calculated by the formula

\begin{equation}
\begin{array}{l}
\Delta \rho^{\text{ex}}_{\text{univ}}(s)  =   
-\frac{\displaystyle \Sigma^{ZZ}_{\text{ex}}(s)-\Sigma^{ZZ}_{\text{ex}}(z)}
{\displaystyle  s-z} 
+\frac{\displaystyle \Sigma^{ZZ}_{\text{ex}}(z)}{\displaystyle z}\\
-\frac{\displaystyle \Sigma^{WW}_{\text{ex}}(0)}{\displaystyle w } +
2\frac{\displaystyle s_W}{\displaystyle c_W}
\frac{\displaystyle \Sigma^{\gamma Z}_{\text{ex}}(0)}
{\displaystyle  z}
\end{array}
\label{deltarho-def}
\end{equation}
where $\Sigma^{V_1V_2}(s)$ represents the transverse part of the vacuum polarization amplitude $\Pi^{V_1V_2}_{\mu\nu}(s)$ between
the vector boson $V_1 - V_2$ calculated at scale s, $w(z) = M^2_{W(Z)}$ and
$\Sigma^{\; \prime} = d\Sigma/dq^2$.

The analytical results for the contributions of the excited lepton doublets to the transverse part of the vector boson self energies were calculated in \cite{GonzalezGarcia:1996nz} using dimensional regularization techniques. In this work, we have calculated the contributions to the vector boson self energies coming from the excited lepton triplet states using mathematica package FeynCalc~\cite{Mertig:1995aa,Shtabovenko:2016aa} and cross-checked our results with direct analytical calculations. The generic formulae for the transverse part of the vector boson self energies for the excited lepton triplet are same as for the case of excited lepton doublets. The transverse part of the vector boson self energy diagram with ordinary lepton and excited lepton in the loop is given by (see \cite{GonzalezGarcia:1996nz} for details)
\begin{equation}
\begin{array}{ll}
\Sigma^{V_1 V_2}_{Ff} =&  
 \frac{\displaystyle 1}{\displaystyle 12 \pi^2} C_{V_1Ff} C_{V_2Ff}
\;
\Biggl\{ 6 q^2 \Lambda^2 + 
q^4 \log\frac{\displaystyle \Lambda^2}{\displaystyle M^2} \\
&- 2 q^2 M^2 -\frac{\displaystyle q^4}{\displaystyle 3} 
\\&+ M^2(2M^2-q^2) + (M^2 - q^2)(2M^2 + q^2)\\
&\times 
\left[-2 +
\left(1-\frac{\displaystyle M^2}{\displaystyle q^2}\right)
\log\left(1-\frac{\displaystyle q^2}{\displaystyle M^2} \right)
\right]
\Biggr\}
\end{array}
\label{pi:eu:final}
\end{equation}
where $V_{1(2)}$ represent the initial (final) vector boson, $q$ is external momentum and $M$ is the mass of excited leptons in the loop. The transverse part of the self energy diagrams that contain two de-generate excited leptons in the loop can be written as

\begin{equation}
\begin{array}{ll}
\Sigma^{V_1 V_2}_{FF} =&  
\frac{\displaystyle 1}{\displaystyle 24\pi^2}  \; 
\frac{\displaystyle q^2}{\displaystyle M^2}
\Biggl\{6 K_{V_1FF} K_{V_2FF} M^2 \Lambda^2 \\
&+ \Bigl[ 2 A_{V_1FF} A_{V_2FF} 
\\ 
&  + 6 \left(A_{V_1FF} K_{V_2FF} + A_{V_2FF} K_{V_1FF}\right) M 
\\&+ 3 K_{V_1FF} K_{V_2FF} 
\left(\frac{\displaystyle q^2}{\displaystyle 3} + 2 M^2\right)
\Bigr ] M^2 
\log \frac{\displaystyle \Lambda^2}{\displaystyle M^2}\\
&+ 4 A_{V_1FF} A_{V_2FF}M^2
\left( \frac{\displaystyle 1}{\displaystyle 3}+ 
\frac{\displaystyle 2 M^2}{\displaystyle q^2}\right)\\&
+ 6 (A_{V_1FF} K_{V_2FF} + A_{V_2FF}
K_{V_1FF}) M^3  \\
&  + K_{V_1FF} K_{V_2FF} M^2
\left(\frac{\displaystyle 5 q^2}{\displaystyle 3} + 4 M^2\right) \\
&- 2 \frac{\displaystyle (4 M^2 -q^2)^{1/2}}{\displaystyle q}
\arctan\left[\frac{\displaystyle q}{\displaystyle (4 M^2 - q^2)^{1/2}}\right]\\&
\times\,\left[
2 A_{V_1FF} A_{V_2FF}M^2
\left( 1+ 
\frac{\displaystyle 2 M^2}{\displaystyle q^2}\right)
\right. \\
&\phantom{\times\,} \left. + 6 (A_{V_1FF} K_{V_2FF} + A_{V_2FF}
K_{V_1FF}) M^3 \right.\\&\phantom{\times\,}
\biggl. + K_{V_1FF} K_{V_2FF} M^2
\left(q^2+ 8 M^2\right) 
\biggr]
\Biggr\}
\end{array}
\label{pi:ee:final}
\end{equation}

In the large-$M$ limit, for $R_Q \equiv q^2/M^2 \ll 1$, we get
\begin{widetext}
\begin{equation}
\begin{array}{ll}
\Sigma^{V_1V_2}_{\text{ex}} &=  \Sigma^{V_1V_2}_{\text{Ff}}+
\Sigma^{V_1V_2}_{\text{FF}} \\
& = 
\frac{\displaystyle M^2}{\displaystyle 12\pi^2} R_Q 
\Biggl\{3\Lambda^2 ( 2C_{V_1Ff} C_{V_2Ff}+K_{V_1FF} K_{V_2FF})
-A_{V_1FF} A_{V_2FF}\\
&  -3 (A_{V_1FF} K_{V_2FF} + A_{V_2FF} K_{V_1FF}) M 
- 6 K_{V_1FF} K_{V_2FF}M^2 - 3 C_{V_1Ff} C_{V_2Ff}M^2
 \\
& + \left[ A_{V_1FF} A_{V_2FF} + 3 (A_{V_1FF} K_{V_2FF} + 
A_{V_2FF} K_{V_1FF}) M 
+ 3 K_{V_1FF} K_{V_2FF} M^2\right] 
\log \frac{\displaystyle \Lambda^2}{\displaystyle M^2}\Biggr\}
\end{array}
\end{equation} 
\end{widetext}

The universal corrections to $\Delta \rho$ coming from the weak iso-spin doublet in the large-$M$ limit are then given by
\begin{eqnarray}
\label{del-rho-doublet} 
\Delta \rho (M_{Z}^{2}) &=&\frac{\alpha R_{z}}{720\pi }\frac{%
C_{W}^{4}+S_{W}^{4}}{C_{W}^{2}S_{W}^{2}}%
\biggl[-20-60 k \sqrt{R_L}-50f^{2}R_{L}  \nonumber \biggr.\\ &&\biggl.-15k^{2}R_{L}+30f^{2}R_{L}\rm{log} R_{L}+15k^{2}R_{L} \rm{log} R_{L}\biggr] \nonumber \\ 
\end{eqnarray}
where $R_Z=M_Z^2/M^2$, $R_L=M^2/\Lambda^2$ and for simplicity we have assumed that $f_1 = f_2=f$ and $k_1=k_2=k$. The universal corrections resulting from weak iso-spin triplet calculated using the couplings given in \eqref{AV},  \eqref{KV} and \eqref{CV}, are given below,
\begin{eqnarray}
\Delta \rho (M_{Z}^{2}) &=&-\frac{\alpha }{72\pi }%
R_{z}\biggl[\biggl(40f^{2}R_{L}+24f^{2}R_{L}\ln R_{L}\nonumber \\ &&+48R_{Z}R_{L}-2k^{2}R_{L}+6R_{L}\ln R_{L}\biggr)\cot ^{2}\theta _{W}
\nonumber\\
&&-(3k^{2}R_{L}+9R_{L}\ln R_{L})\tan
^{2}\theta _{W}\biggr]
\label{del-rho-triplet}
\end{eqnarray}

The results presented in \eqref{del-rho-doublet} and  \eqref{del-rho-triplet} were obtained with the assumption that the masses of the excited leptons are degenerate. However, mass splitting between the excited leptons can have considerable impact on the predictions for $\Delta \rho$. For example, the $W$ boson self energy $\Sigma_{ex}^{WW}(0)=0$  if the excited leptons in the loop have equal masses. However there will be non-zero contribution from $W$ boson self energy $\Sigma_{ex}^{WW}(0)$ to $\Delta \rho$ if the masses of the excited leptons in the loop are different. These contributions are given by the formula  
\begin{eqnarray}
\Delta \rho  &=&-\frac{G_{\mu } (a_1^2) }{2\sqrt{2}\pi ^{2}}%
\biggl\{F(M_{E^{0}}^{2},M_{E^{-}}^{2})+F(M_{E^{-}}^{2},M_{E^{--}}^{2}) \biggr.\nonumber\\
&&-2(M_{E^{0}}-M_{E^{-}})^{2}\biggl[\ln\frac{\Lambda^{2}}{M_{E^{0}}M_{E^{-}}}\nonumber\\&&%
\phantom{xxxxxxx}-\frac{M_{E^{0}}^{2}+M_{E^{-}}^{2}}{2(M_{E^{0}}^{2}-M_{E^{-}}^{2})}\ln\frac{%
M_{E^{0}}^{2}}{M_{E^{-}}^{2}}\biggr] \nonumber \\
&&-2(M_{E^{-}}-M_{E^{--}})^{2}\biggl[\ln\frac{\Lambda ^{2}}{M_{E^{-}}M_{E^{--}}}\nonumber\\&&\phantom{xxxxxxx}-%
\frac{M_{E^{-}}^{2}+M_{E^{--}}^{2}}{2(M_{E^{-}}^{2}-M_{E^{--}}^{2})}\ln\frac{%
M_{E^{-}}^{2}}{M_{E^{--}}^{2}}\biggr]\biggr\}
\label{del-rho-mass-split}
\end{eqnarray}
Where $G_{\mu }$ is the usual Fermi constant. The function $F_{0}(m_{1}^{2},m_{2}^{2})$ which is proportional to the mass splitting is defined as

\begin{equation}
F_{0}(m_{1}^{2},m_{2}^{2})=m_{1}^{2}+m_{2}^{2}-\frac{2m_{1}^{2}m_{2}^{2}}{m_{1}^{2}-m_{2}^{2}}\ln 
\frac{m_{1}^{2}}{m_{2}^{2}}  \label{ffunction}
\end{equation}

\section{Numerical Results}
\label{Nresults}
The new physics effects to the electroweak precision observables are usually described in terms of a set of three independent parameters called oblique parameters $S$, $T$ and $U$ \cite{Peskin:1990zt, Peskin:1991sw}. The $\Delta\rho$ is related to the oblique parameter T via the relation 
\begin{equation}
\Delta \rho= \hat{\alpha}(M_Z) T 
\end{equation}
where electromagnetic fine structure constant $\hat{\alpha}(M_Z)^{-1}=127.950\pm0.017$~\cite{Tanabashi:2018oca}. The current value for $T=0.07\pm0.12$ \cite{Tanabashi:2018oca} can be used to put bounds on the value of $\Delta\rho$ given by 
\begin{equation}
-0.00039 \leq \Delta\rho \leq 0.00148
\label{Exp-bound-delrho}
\end{equation} 
%\subsection{Degenerate Masses}

\begin{figure}[htb!]
\begin{center}
\includegraphics[scale=0.85]{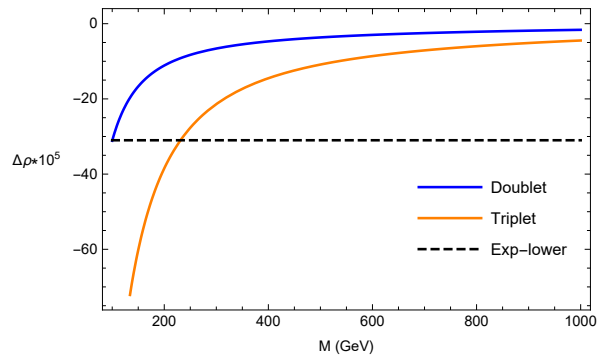}
\end{center}
\caption{Excited lepton doublet (blue line) and triplet (orange line) contributions to $\Delta \rho $ as a function of excited lepton mass $M$. Dashed black line shows the experimental lower bound on $\Delta\rho$.}
\label{fig:degen}
\end{figure}

Using the formulae given in \eqref{del-rho-doublet} and  \eqref{del-rho-triplet}, we can estimate the contributions of the excited leptons to the $\Delta \rho$. In Fig.~\ref{fig:degen} we show our numerical results for excited lepton doublet and triplet contributions to $\Delta\rho$ as a function of excited lepton mass $M$. Here we fix the cut-off scale $\Lambda=1$ TeV.  As one can see the doublet (solid blue line) and triplet (solid orange line) contributions in the degenerate case are well within the experimental bounds on  $\Delta\rho$ given in  \eqref{Exp-bound-delrho} except for a small range of excited lepton mass  $M<200$ GeV which however is already excluded by the experimental searches.  

 We assume that main contribution to the excited fermion masses arise at scale above EWSB and therefore we assume that they are of similar magnitude with small differences arising from SU(2) breaking contributions like couplings to exotic Higgs bosons. As already discussed, this computation is very sensitive to even small mass splittings of the SU(2) multiplets and can result in large contributions to $\Delta\rho$ in the form of $W$ boson self energy as shown in \eqref{del-rho-mass-split}. Therefore we introduce mass splitting between the excited Fermions and study its impact on the $\Delta \rho$ for the doublet and triplet contributions. Here we define $M_{E^{0}}=M-\delta M$, $M_{E^{-}}=M$, and $M_{E^{--}}=M+ \delta M$.

\begin{figure}[htb!]
\begin{center}
\includegraphics[scale=0.8]{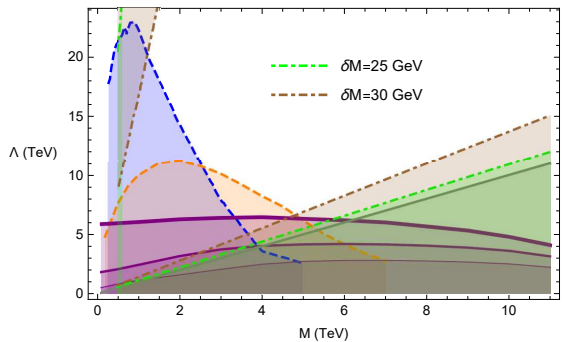}
\end{center}
\caption{$\Delta \rho$ in $(M, \Lambda)$ plane with $\delta M= 25$ GeV (dot-dashed green lines) and $\delta M= 30$ GeV (dot-dashed brown lines) for the case of excited lepton doublet. $M=\Lambda$ line (solid gray), unitarity bound (purple lines)\cite{Biondini:2019tcc} and exclusion limits (blue and orange dashed lines) \cite{Sirunyan:2018zzr,CMS-PAS-EXO-18-013} from run 2 for charged leptons searches with two different final states are also shown for comparison.}
\label{fig:DoubletPlot}
\end{figure}

Extending our analysis for the case of non degenerate masses, we show the contours of $\Delta \rho$ in $(M,\Lambda)$ plane in Fig.~\ref{fig:DoubletPlot}. We find that constrained areas by $\Delta\rho$ above the line $\Lambda=M$ (solid grey line) can be found for mass splittings larger than about 20 GeV, excluded areas  with  $\delta M= 25$ GeV (Dot-dashed green line) and $\delta M= 30$ GeV (Dot-dashed brown line) are shown. Area below the lower dot-dashed lines is excluded due to the lower bound on $\Delta\rho$ while the area above the upper dot-dashed lines is excluded by upper bound on $\Delta\rho$. We have significant doublet contributions for $\delta M= 25$ GeV (green dot-dash line)  which can result in exclusion of small area in the lower right corner of the $(M,\Lambda)$ plane due to the lower experimental bound on $\Delta\rho$. 
Increasing the $\delta M$ to 30 GeV not only results in the exclusion of larger part of the  $(M,\Lambda)$ plane in the lower right corner of the plot but some area in the upper left corner also gets excluded, as indicated by the dot-dashed brown line, due to upper experimental limit on $\Delta\rho$. We also show the experimental findings here depicting the exclusion regions in the $(M,\Lambda)$ plane with 95$\%$ C.L. \cite{Sirunyan:2018zzr,CMS-PAS-EXO-18-013}. The dashed blue line corresponds to the recent CMS search in $\ell \ell \gamma$ channel with integrated luminosity of $35.9$  $\rm{fb^{-1}}$ and orange line  corresponds to the search for excited lepton decaying to two electrons or two muon and two jets via contact interaction with a total integrated luminosity of $77.4$ $\rm{fb^{-1}}$. The perturbative unitarity bounds for a composite fermion model were studied in \cite{Biondini:2019tcc}. These bounds are depicted using purple line with decreasing thickness corresponding to 100$\%$, 95$\%$ and 50$\%$ event fractions respectively that satisfy unitarity bounds. The shaded area is excluded due to unitarity bounds (purple lines) and direct searches at the LHC (blue and orange lines).

\begin{figure}[htb!]
\begin{center}
\includegraphics[scale=0.8]{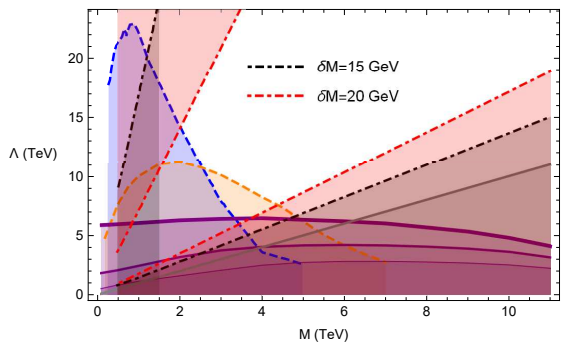}
\end{center}
\caption{$\Delta \rho$ in $(M, \Lambda)$ plane with $\delta M= 15$ GeV (dot-dashed black line) and $\delta M= 20$ GeV (dot-dashed red lines) for the case of excited lepton triplet.  The $(M = \Lambda)$ line (solid grey), unitarity bound (purple lines) \cite{Biondini:2019tcc} and exclusion limits (blue and orange dashed lines) \cite{Sirunyan:2018zzr, CMS-PAS-EXO-18-013} from run 2 for charged leptons searches with two different final states are also shown for comparison.}
\label{fig:TripletPlot}
\end{figure}

In Fig~\ref{fig:TripletPlot}, we show the $\Delta \rho$ in $(M, \Lambda)$ plane for the case of excited lepton triplet. In this case the space above the line $\Lambda=M$ (solid grey line) is constrained by $\Delta \rho$ for mass splittings larger than 10 GeV. Here the space above  the bounds are more stringent compared to unitarity bounds even for $\delta M= 15$ GeV as indicated by the dot-dash black line. Increasing the mass splitting between the triplets to $\delta M= 20$ GeV result in the exclusion of larger area in the lower right region (area below the lower red line) as well as the upper left corner of the plot (area above the upper red line). 

\section{Conclusions}
\label{sec:conclusions}
The phenomenology of the effective interactions of a composite scenario of the standard model fermions based  on extended isospin multiplets ($I_W=1, 3/2$) has recently been revived with several studies of the  production at LHC of the corresponding states with exotic charge such as doubly charged leptons and quarks of charge $Q=5/3 e$.   
On the other end  both the ATLAS and CMS Collaborations have been  testing this model  at increasingly higher energies and luminosities extending the corresponding bounds in the parameter space $[\Lambda,M]$  with new analyses.  Plans are laid out to further these studies at the high luminosity (HL)  option of the LHC collider.

In addition to direct searches for the excited fermions at the LHC  also perturbative unitarity bounds proved very effective in constraining the composite fermion models~\cite{Biondini:2019tcc}. In this paper, however, we have adopted yet another approach in order to constrain the excited fermion paramter space. We analyzed the effects of excited leptons to the electroweak precision observables (EWPO) in particular the $\Delta \rho$ parameter. As a first step, we calculated the couplings of the $I_W=1$ multiplet excited leptons to the ordinary leptons and gauge bosons and the excited lepton coupling to the excited lepton and gauge bosons. In the second step, we used these couplings to calculate the one-loop contributions to the $W$, $Z$ and $\gamma$ self-energies for the case of the lepton triplet. These self-energy contributions were then used to estimate the value of $\Delta \rho$. 
For the case of mass degenerate excited leptons, their contributions to $\Delta \rho$ turned out to be very small. However, mass splitting between the excited leptons can result in the large contributions to the $\Delta \rho$ coming from the $W$ boson self energy diagram. These contributions are directly proportional to the mass splitting and can results in the exclusion of significant regions in the [$\Lambda,M$] plane compared to the present experimental bounds and those coming from perturbative unitarity. We believe that this study will prove helpful for the present and future experimental searches on these models.

\begin{acknowledgments}
M.~Rehman  and M.~E. Gomez wish to thank the Istituto Nazionale di Fisica Nucleare, INFN, Sezione di Perugia,  for kind hospitality and  support for  collaboration visits during  the early and final stages of this work. The research of M.E.G. was supported by the Spanish MINECO, under grants FPA2017-86380 and PID2019-107844GB-C22.
\end{acknowledgments}

\bibliography{main}% Produces the bibliography via BibTeX.

\end{document}